# High resolution imaging of reconstructed domains and moiré patterns in functional van der Waals heterostructure devices

Andrey Sushko[1,†], Kristiaan De Greve[1,5,†], Trond I. Andersen[1,†], Giovanni Scuri[1,†], You Zhou[1,2], Jiho Sung[1], Kenji Watanabe[3], Takashi Taniguchi[3], Philip Kim[1,4], Hongkun Park[1,2,*], Mikhail D. Lukin[1,*]

[1]*Department of Physics, Harvard University, Cambridge, MA 02138, USA*

[2]*Department of Chemistry and Chemical Biology, Harvard University, Cambridge, MA 02138, USA*

[3]*Advanced Materials Laboratory, National Institute for Materials Science, 1-1 Namiki, Tsukuba, 305-0044, Japan*

[4]*John A. Paulson School of Engineering and Applied Sciences, Harvard University, Cambridge, MA 02138, USA*

[5]*Currently at imec, 3001 Leuven, Belgium*

[†] *These authors contributed equally to this work*

[*] *To whom correspondence should be addressed: lukin@physics.harvard.edu, hongkun_park@harvard.edu*

**The optical and electronic properties of van der Waals (vdW) heterostructures depend strongly on the atomic stacking order of the constituent layers. This is exemplified by periodic variation of the local atomic registry, known as moiré patterns, giving rise to superconductivity and ferromagnetism in twisted bilayer graphene and novel exciton states in transition metal dichalcogenides (TMD) heterobilayers. However, the presence of the nanometer-scale moiré superlattices is typically deduced indirectly, because conventional imaging techniques, such as transmission electron microscopy (TEM), require special sample preparation that is incompatible with most optical and transport measurements. Here, we demonstrate a method that uses a secondary electron microscope to directly image the local stacking order in fully hexagonal boron nitride (hBN) encapsulated, gated vdW heterostructure devices on standard Si-substrates. Using this method, we demonstrate imaging of reconstructed moiré patterns in stacked TMDs, ABC/ABA stacking order in graphene**

**multilayers, and AB/BA boundaries in bilayer graphene. Furthermore, we show that the technique is non-destructive, thus unlocking the possibility of directly correlating local stacking order with optical and electronic properties, crucial to the development of vdW heterostructure devices with precisely controlled functionality.**

Characterizing stacking orders, stacking defects and moiré superlattices is essential to a complete understanding of and control over 2D material systems. Studies of effects such as Mott insulator and superconducting states in ABC-stacked trilayer graphene[1, 2], and topologically protected states along the AB/BA boundaries in bilayer graphene[3–6] require spatially-resolved determination of the stacking order. Moreover, in twisted vdW structures, imaging of the moiré pattern is essential for understanding how optical and electronic properties depend on moiré periodicity[7–10], as well as effects of superlattice inhomogeneity and lattice reconstruction. In fact, the layers can twist and reconstruct during fabrication, often causing the moiré superlattice to vary spatially and differ from the target stacking angle, greatly complicating interpretation of experiments. For instance, it was recently shown that interactions between the lattices in a bilayer graphene structure caused spatially varying reconstruction patterns that differed significantly from the conventional moiré picture of a smoothly-varying potential[11].

Direct imaging of the lattice structure is typically achieved with TEM[11, 12], which requires placing the samples on thin (50 nm) TEM grids – incompatible with most device fabrication techniques. In the case of post-measurement TEM imaging, the 2D heterostructures must be separated from the contacts, picked up and transferred onto the TEM grid. Apart from being technically challenging, this procedure likely distorts the lattice alignment. Moiré patterns have also been imaged using atomic force microscopy (AFM)[13] and scanning tunneling spectroscopy (STS)[14], but this requires direct contact with the active area of the heterostructure, thus preventing the use of hBN encapsulation and top gates. While scanning electron microscopy (SEM) techniques do not generally suffer from these limitations, conventional SEM techniques used for crystallographic imaging, such as angle-resolved electron backscatter diffraction (EBSD)[15] and electron channeling contrast imaging (ECCI)[16] rely on detection of backscattered primary electrons, which is not efficient enough to probe mono- or bilayer materials. In recent studies of conventional, covalently bonded semiconductors, it was pointed out that the channeling of primary electrons through a lattice also affects the generation of secondary electrons[17] allowing for direct imaging of the surface lattice ordering of SiC.

Our approach, illustrated in Fig. 1a, involves the use of an SEM with a generic, yet highly sensitive Everhart-Thornley (E-T) secondary electron detector[18]. The device under study is positioned at an angle with respect to the incoming electron beam - the energy of which is varied between 0.5 keV and 15 keV. The SEM signal comes from the generation of secondary electrons at different depths, the extent of which depends on the ability of the primary beam to channel through the stack. In order to understand how the technique, which we here refer to as "channeling modulated secondary electron imaging", resolves atomic stacking order, we first consider a simple system consisting of naturally occurring AB/BA stacking orders in graphene.

Figures 1b-c show images of a bilayer graphene flake on a $SiO_2$-on-Si substrate, acquired with optical microscopy and SEM, respectively. While the former is homogeneous, confirming that the flake has uniform thickness, SEM imaging at a $24°$ tilt angle uncovers two types of distinct domains. As the stage is rotated azimuthally, the number of secondary electrons collected from these domains exhibits $120°$-periodic oscillations, with a $60°$ phase shift between the two domain types (Fig. 1d). We attribute the observed domains to different levels of electron channeling at a given azimuthal angle through naturally occurring regions of AB- and BA-stacked graphene[3–5,19]. Such channeling arises at particular polar and azimuthal angles, when the incoming electrons are oriented parallel to the open cavities or "channels" within the atomic lattice, allowing them to travel through the material with minimal scattering. In particular, channeling conditions are optimized every $120°$ (Fig. 1d, inset), and the $60°$ relative angle between AB and BA domains causes the observed phase difference. Since graphene has a lower atomic number, Z, than the substrate, it generates fewer secondary electrons (Fig. 1c, inset)[20]. Thus, for graphene on oxide substrates, the secondary yield

maxima occur when channeling is maximized.

Turning to a thicker (approx. 6 layer) graphene flake, which also exhibits domains in SEM imaging that are not observable optically (Figs. 2a-b), we find a more complex azimuthal dependence (Fig. 2c). While one domain type (red) is $60°$-periodic in azimuth angle, the other (blue) exhibits a period of $120°$, reflecting a lower inherent symmetry. In particular, the two curves have nearly equal amplitude at $60°$ but the domain represented in blue exhibits significantly stronger signal at $0°$ and $120°$, suggesting a greater degree of channeling at these angles. To highlight the lower symmetry regions, we present in Fig. 2d a spatial map of the ratio between the signals at $0°$ and $60°$ as a function of position - clearly pronouncing the middle section of the flake. We find that this more complex pattern is due to ABA- and ABC-stacking orders[21], as can be independently verified with Raman spectroscopy[22] through the position of the Raman 2D peak (Fig. 2e). Indeed, consistent with our observations, the two domain types are expected to have similar scattering cross sections at certain rotation angles (Fig. 2c, left inset), while at other angles, the ABC domain allows for "complete channeling" (Fig. 2c, right inset), giving rise to enhanced SEM signal every $120°$.

The graphene domain imaging demonstrated above is applicable to twisted van der Waals heterostructures. Figure 3a shows an SEM image of an hBN-encapsulated, dual-gated, twisted bilayer of WSe$_2$ (Fig. 3a, inset). In the SEM image, clear triangular lattice reconstructions are visible, with domain sizes ranging from ∼50 nm to ∼300 nm. These results indicate that the relative twist angle is strongly position dependent, and that interactions between the respective lattices are strong enough to dominate over the in-plane stiffness of the respective layers. The end result is that the two WSe$_2$ layers lock locally into energetically favourable AB or BA stacking configurations - similar to what was previously observed in twisted bilayer graphene through TEM imaging[11]. Critically, our imaging technique is compatible with optoelectronic device operation, retaining narrow ∼3 nm linewidths, and electrostatic gate tunability after over 6 hrs of SEM imaging at energies up to 15 keV (see SI). These observations indicate that the device maintains excellent optical properties after SEM imaging, and thus demonstrate the powerful potential of the present technique: namely, spatially correlating optical spectra with directly imaged local lattice configuration.

A key feature of our technique is the ability to image 2D materials fully encapsulated in hBN, which is known to enhance both optical and electronic properties[23]. In Fig. 4, we explore the dependence of the SEM contrast on hBN thickness, by imaging a heterostructure composed of a stepped $WSe_2$ flake under a perpendicularly stepped hBN flake. An optical image of the sample and the hBN thickness profile (acquired with AFM) are shown in Fig. 4a. As with general secondary electron techniques, the depth to which we can image is determined not by the total penetration of the primary beam (typically on the order of microns) but rather by the ability of secondary electrons to escape the material. Increasing the acceleration voltage increases the energy of the secondary electrons[20], corresponding to a greater imaging depth, as illustrated in Fig. 4b. At primary energies below 1.5 keV (left) the secondary electrons are only collected from the top layers of hBN, with no $WSe_2$ signal through hBN thicker than $\sim$ 10 nm. At an energy of 3 keV (right) the imaging depth exceeds the thickness of the heterostructure, revealing the complete thickness variation in the hBN and underlying $WSe_2$. Increased imaging depth, however, reduces individual layer contrast, as can be seen in the mono-, bi- and tri-layer $WSe_2$ steps along the right edge of the image. Nevertheless, at a reasonable acceleration voltage, monolayer steps in the TMD are clearly visible through more than 27 nm of hBN.

To determine the visibility of different stacking orders through increasingly thick hBN encapsulation, we focus on the azimuthal angle dependence from natural AA'-stacked $WSe_2$. The azimuthal dependence of the bilayer and trilayer SEM signal for different hBN thicknesses and the corresponding contrast (root-mean-square amplitude) are displayed in Figs. 4c,d and e, respectively. Due to the crystalline nature of the top hBN, the raw secondary yield of the bilayer is convolved with channeling effects in the hBN. Yet, we can conveniently correct for this convolution by dividing the raw yield by the secondary yield from an adjacent region of equal hBN thickness but without underlying TMD. We observe that contrast indeed decreases with hBN thickness, but nevertheless persists through to 27 nm. Our ability to resolve signals at large depth is mainly limited by uncertainty in this background deconvolution, as the amplitude of the channeling effect in thick hBN significantly exceeds that of the bilayer. Spatially resolvable contrast, for instance, between AB/BA domains in a bilayer, can of course be imaged directly without such compensation. Since 27 nm thick top-layer hBN is sufficient for most device applications, it appears that our technique is applicable to a wide range of encapsulated vdW heterostructures.

Our observations can be understood from a simple, semi-classical model of channeling physics. By representing the 3D configuration of atoms within the material, and assigning a soft radially-dependent electron transmission probability[24] to each, we compute the mean transmission for a series of parallel electron beams at a given polar and azimuthal angle. For a TMD bilayer encapsulated in lighter elements with lower secondary yield (such as the device in Fig. 3) contrast mainly results from the difference in direct secondary generation in the bilayer. Hence, it is sufficient to model the secondary emission as proportional to the scattering probability in the bilayer in order to reconstruct an azimuthal dependence that closely resembles the experimental data (Fig. 3c). In this instance, a more complex azimuthal dependence than in Fig. 1d arises due to the 3-dimensional structure of $WSe_2$ along with a shallower, $40°$ polar angle. An important difference between TMDs and few-layer graphene (Figs. 1,2), is that the latter has a lower secondary yield than the $SiO_2$ substrate and therefore produces a "negative" signal, in the sense that enhanced channeling through the top layer and into the substrate increases overall detected secondary yield. In this scenario, we model the yield as proportional to the transmission probability and accurately recover the dominant features of ABA/ABC stacked few-layer graphene (Fig. 2c). Higher order features in between the dominant peaks are highly dependent on the particular scattering parameters used, as well as secondary emission from the graphite itself and are therefore not well represented in this simplified model.

Finally, we note a method to resolve surface-level structure by probing the material with a low energy beam ( 500 eV). Due to the rapid loss of secondary yield with increasing depth, the stacking orientation of the top two layers of bulk graphite can be readily measured (see SI), allowing for analysis of low-Z materials even over arbitrarily thick low-Z substrate, such as bilayer graphene over thick hBN.

The stacking method by which vdW heterostructures can be assembled[23, 25, 26] allows for control and tuning of the stacking order in generic devices, where it can now be used as a versatile control knob. Recently, moiré superlattices have been shown to cause novel and exotic phases in transport and optical spectra, including the Hofstadter butterfly fractal pattern in the Quantum Hall spectrum of graphene-hBN superlattices[27–29], exciton trapping in TMD heterobilayers[30–33] and unconventional superconductivity and ferromagnetism in twisted graphene bilayers [7, 8, 34]. However, in most of these experiments, the presence of the moiré superlattices was not directly imaged, but rather inferred indirectly, either from far-field optical and transport measurements or from transmission electron microscopy (TEM) of other samples. Our observations demonstrate that channeling modulated secondary electron imaging can be used to directly image the atomic stacking orders of vdW heterostructures, including bilayer and multilayer graphene, and twisted bilayer TMDs. In all cases studied, simple computational modeling qualitatively predicts imaging contrast, enabling the systematic application of this technique to other materials and device structures. Compatible with complete, encapsulated and gated devices on standard substrates, our technique allows for direct correlation of atomic stacking order with optical and transport properties, as already explored in a complimentary study involving exciton arrays[35]. For these reasons, channeling modulated secondary electron imaging can be indispensable for the further development of novel electronic and optical devices, including the realization of exotic emitter arrays, solid state quantum simulation platforms[36, 37], and new quantum optical systems[38].

**Acknowledgements** We thank Bernhard Urbaszek for valuable review and feedback on the manuscript, and Hyobin Yoo, Dominik Wild, Andrew Joe, Johannes Knoerzer and Wilfried Vandervorst for continued discussions, feedback and suggestions.

**Funding** We acknowledge support from the DoD Vannevar Bush Faculty Fellowship (N00014-16-1-2825 for H.P.,

N00014-18-1-2877 for P.K.), NSF (PHY-1506284 for H.P. and M.D.L.), NSF CUA (PHY-1125846 for H.P. and M.D.L.), AFOSR MURI (FA9550-17-1-0002), ARL (W911NF1520067 for H.P. and M.D.L.), the Gordon and Betty Moore Foundation (GBMF4543 for P.K.), ONR MURI (N00014-15-1-2761 for P.K.), and Samsung Electronics (for P.K. and H.P.). All fabrication was performed at the Center for Nanoscale Systems (CNS), a member of the National Nanotechnology Coordinated Infrastructure Network (NNCI), which is supported by the National Science Foundation under NSF award 1541959. K.W. and T.T. acknowledge support from the Elemental Strategy Initiative conducted by the MEXT, Japan and the CREST (JPMJCR15F3), JST. A.S. acknowledges support from the Fannie and John Hertz Foundation and the Paul and Daisy Soros Fellowships for New Americans.

**Competing Interests**  The authors declare that they have no competing interests.

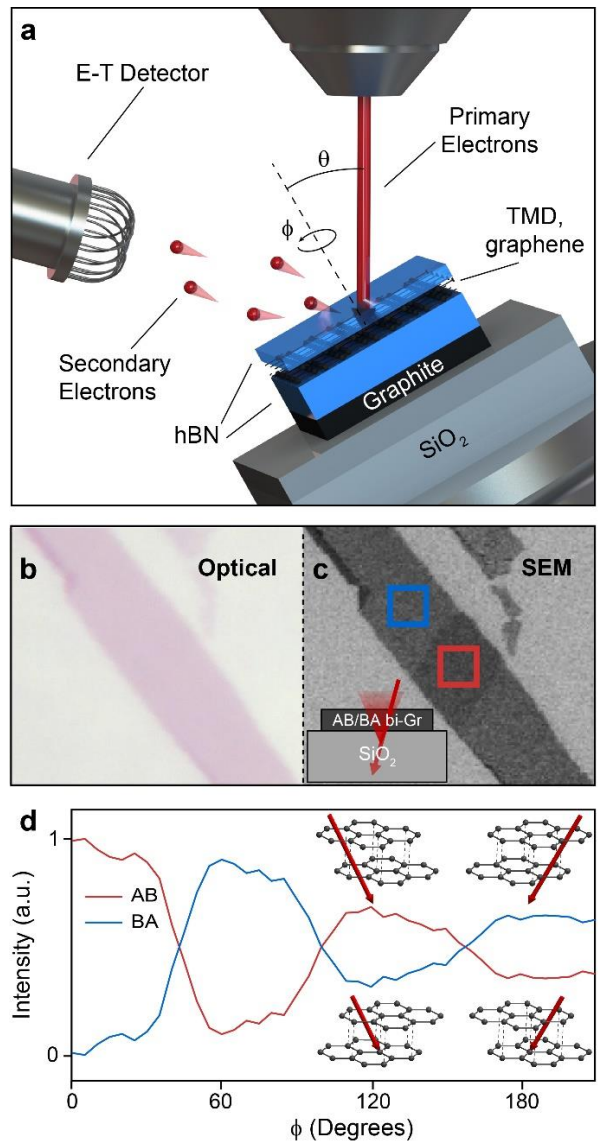

**Figure 1:** | **SEM based imaging of vdW heterostructures: operating principle and simple example a**) Schematic of operating principle: an accelerated electron beam is focused onto the device, which is placed at a non-zero polar angle ($\theta$) and can also be rotated azimuthally ($\varphi$). Channeling of the incident electrons through the van der Waals stack and subsequent secondary electron generation depend on the effective scattering cross-section at the given incidence angle. Collecting the secondary electrons on a sensitive Everhart-Thornley (E-T) detector thus allows for imaging the local stacking order. **b**) Optical image of a naturally occurring bilayer graphene flake, showing homogeneous thickness. **c**) SEM image of the same flake as in (**b**) ($\theta = 24°$, $\varphi = 5°$), revealing areas of different contrast. **d**) Dependence of SEM secondary yield on azimuthal angle for locations marked by boxes in (**b**) (normalized to the mean). The two regions display $120°$ symmetry and $180°$ phase offset, consistent with AB and BA domains: every $120°$, channeling is maximized for one domain type, and minimized for the other (inset schematics). Contrast decays at increasing angle due to gradual charging of the substrate over time.

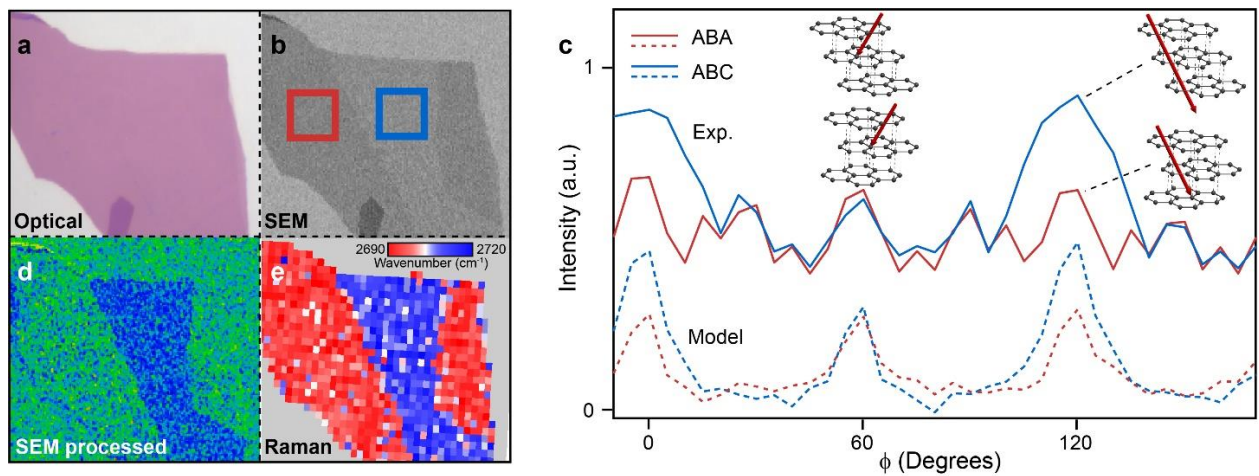

**Figure 2:** | **Channeling contrast based SEM imaging of multilayer graphene a-b**) Optical and SEM images of a natural few-layer graphene flake, again highlighting structural contrast that is not visible optically. **c**) Azimuthal angle dependence, showing qualitatively different behavior in the two boxed regions in (**b**), consistent with ABC and ABA stacking: while ABC stacking exhibits fully open channels every $120°$, ABA stacking shows only partially open channels every $60°$. **d**) Difference between two SEM images with a $60°$ offset in azimuthal angle. As expected from (**c**), this selectively reveals ABC domains. **e**) Spatial map of Raman 2D-peak position, confirming the ABC/ABA stacking order in the different domains.

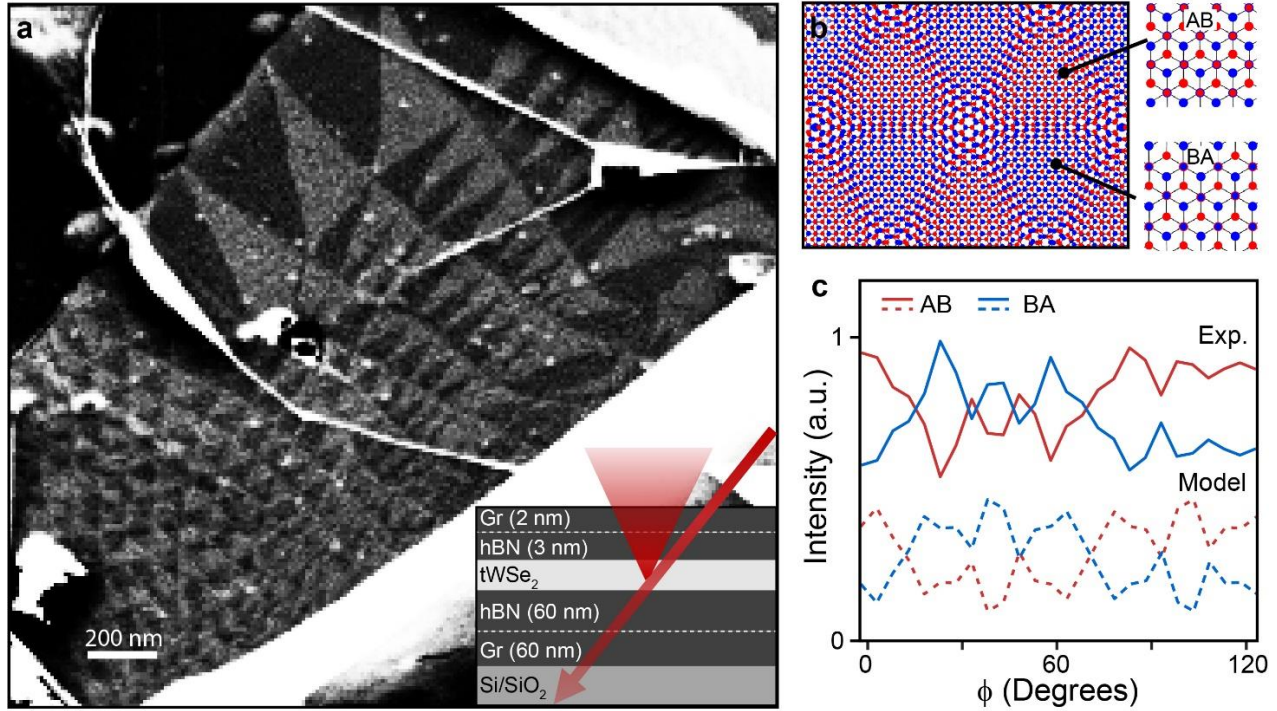

**Figure 3:** | **Imaging of fully hBN-encapsulated twisted vdW heterostructure a**) Representative images of the contrast that appears in a twisted bilayer $WSe_2$ device. Due to the interaction between the respective layers, a spatially varying reconstruction pattern develops, determined by strain and the local mismatch angle (see text for details). Inset: Device schematic. **b**) Schematic illustration of a twisted bilayer in the strained reconstruction regime. For small misalignment angles, the interlayer interactions locally lock the lattice into domains of AB and BA stacking with strain accumulated at layer boundaries. For a given polar angle, the two stacking orders facilitate optimal channeling conditions at different azimuth angles, giving rise to a contrast image. **c**) Azimuthal angle dependence of the signal from AB and BA domains (solid red and blue curves, respectively), along with predictions based on the model presented in the main text.

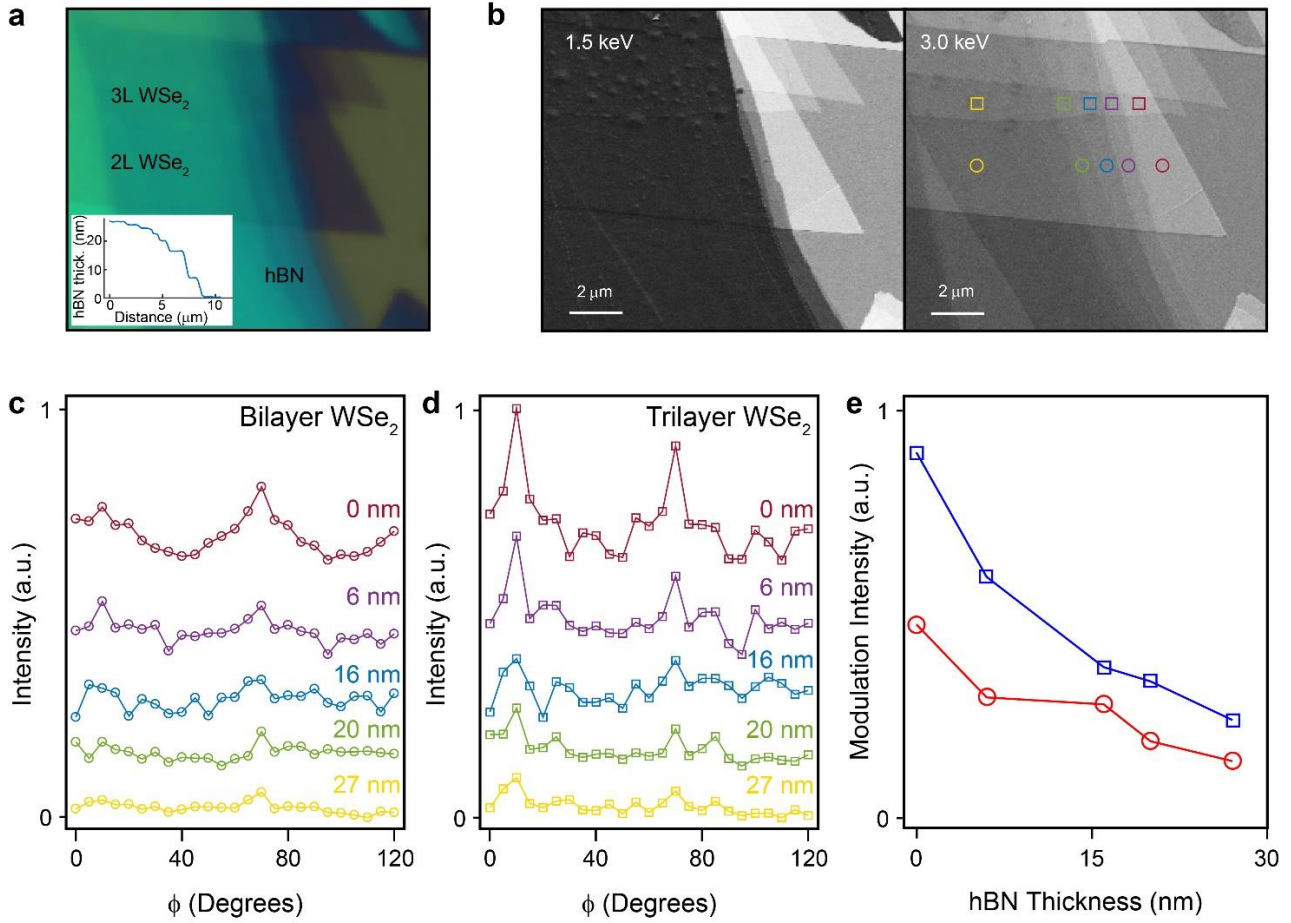

**Figure 4:** | **Depth dependence of SEM contrast a**) Optical image of a naturally stepped WSe$_2$ flake stacked underneath a stepped hBN flake. Inset: AFM of stepped hBN thickness. **b**) SEM images of the sample in (**a**) at 1.5 keV (left) and 3 keV (right) primary acceleration voltage. Tuning the primary voltage alters the secondary electron energy spectrum, allowing for greater escape depth at higher acceleration voltage. For 1.5 keV, the escape depth is ∼10 nm, restricting collected secondary signal to the hBN. At 3 keV, the depth exceeds 30 nm, revealing the underlying stepped WSe$_2$ structure. **c-d**) Azimuthal angle dependence of secondary yield from bilayer (**c**) and trilayer (**d**) WSe$_2$ under varying thickness of hBN. The curves are normalized by dividing by the yield of bare silicon oxide underneath an equal thickness of hBN. **e**) RMS amplitude of the modulation signal for bi- and tri-layer WSe$_2$ (red and blue squares, respectively) as a function of encapsulating layer thickness. Our data indicate that domain contrast in bilayer TMDs is resolvable even through 20 nm of hBN.

# Supplementary material for: High resolution imaging of reconstructed domains and moiré patterns in functional van der Waals heterostructure devices

Andrey Sushko[1,†], Kristiaan De Greve[1,5,†], Trond I. Andersen[1,†], Giovanni Scuri[1,†], You Zhou[1,2], Jiho Sung[1], Kenji Watanabe[3], Takashi Taniguchi[3], Philip Kim[1,4], Hongkun Park[1,2,*], Mikhail D. Lukin[1,*]

[1]*Department of Physics, Harvard University, Cambridge, MA 02138, USA*

[2]*Department of Chemistry and Chemical Biology, Harvard University, Cambridge, MA 02138, USA*

[3]*Advanced Materials Laboratory, National Institute for Materials Science, 1-1 Namiki, Tsukuba, 305-0044, Japan*

[4]*John A. Paulson School of Engineering and Applied Sciences, Harvard University, Cambridge, MA 02138, USA*

[5]*Currently at imec, 3001 Leuven, Belgium*

[†] *These authors contributed equally to this work*

[*] *To whom correspondence should be addressed: lukin@physics.harvard.edu, hongkun park@harvard.edu*

**Polar angle variation**

The azimuthal angle dependence for graphene in Figs. 1 and 2 is collected at a polar angle of $\theta = 24°$ which approximately corresponds to a displacement of one carbon-carbon bond length per layer. For ABC stacking, this is the steepest angle at which maximal channeling can occur as it corresponds exactly to the inter-layer stacking offset. Owing to the symmetry of the lattice, however, equivalent behavior (with a 60° phase shift in $\varphi$) is obtained for a displacement of two bond-lengths per layer, corresponding to $\theta \sim 41°$. This is demonstrated in Fig. S1, where we image the same graphene flake as in Fig. 2 at these two polar angles ($\theta = 24°, 41°$), using two different azimuthal angles ($\varphi = 0°, 60°$). Under both polar angles, a high degree of channeling in ABC graphene is observed for one of the values of $\varphi$ and not for the other. However, while the ABC region appears at $\varphi = 60°$ for $\theta = 24°$, it appears for $\varphi = 0°$ for $\theta = 41°$.

**Imaging of stepped graphene and TMD flakes**

In order to explore the ability to differentiate stacking orders of graphene on top of materials with similar secondary yield (graphene, hBN), we image thick, stepped graphite flakes (Fig. S2a-c). By using a relatively small acceleration voltage (500 eV), the flake is sufficiently thick that the secondary yield depends on the stacking order of the surface layers, rather than the total thickness. In particular, the SEM signal depends on whether the top two layers are AB- or BA-stacked, and therefore alternates across steps of monotonically changing thickness (Fig. S2a). Similar to in Fig. 1d in the main text, the two stacking orders exhibit 120° contrast oscillations with a 60° offset.

Similar measurements were conducted in a stepped multilayer $MoSe_2$ flake underneath a 10 nm layer of gold (Fig. S2d-f). Again, the SEM contrast depends on the stacking order of the top two layers, and thus alternates with the parity of the number of layers (Fig. S2d). In this case, the angular dependence (Fig. S2f) is more complex than in graphene, due to the 3D nature of the $MoSe_2$ layers, and also because the polar angle used here (30°) is not commensurate with the dominant channels in $MoSe_2$ so does not allow for maximal channeling

**Compatibility with optoelectronic measurements**

Critically, our technique is compatible with optoelectronic device operation, as shown in gate-dependent photoluminescence measurements (Fig. S3) of a hBN-encapsulated monolayer $WSe_2$ device that was imaged with SEM for several hours. We observe clear exciton emission from both bright ($X_0$) and dark ($X_D$) exciton states, with very effective gate-tunability, manifested by the appearance of red-shifted charged excitons ($X^+$ and $X^-$) in the p- and n- doped regimes. The intrinsic regime occurs for gate voltages near 0 V (-3.5 V< $V_G$ <0.5 V) at a hBN dielectric thickness of 30 nm, indicating that the sample is only minorly doped after imaging. Moreover, the device exhibits very narrow (~3 nm) linewidths after imaging, allowing for observing even the small (~2 nm) splitting of inter- and intravalley trions due to exchange interactions. These observations indicate that the device maintains excellent optical properties and effective gate tunability after SEM imaging.

### Signal corrections and hBN background subtraction

When imaging thin materials on an $SiO_2$ substrate for a prolonged period of time, a noticeable degree of charging can be seen in the substrate, gradually altering the level of secondary emission. This charging is compensated for in our data by dividing out the yield from an uncovered region of substrate. For systems with natural contrast, such as AB/BA domains in graphene or twisted $WSe_2$, it is further helpful to normalize to the mean of the two regions so as to eliminate drift due to charging effects and reduce the impact of angle-dependent yield variation due to channeling in the encapsulation layer or substrate. When collecting an azimuthal dependence for materials without domain contrast, such as AA' stacked natural $WSe_2$ bilayers, it is necessary to compensate for channeling effects in surrounding layers carefully. Fig. S4 shows the azimuthal dependence for the bilayer part of the device discussed in Fig. 4, before and after compensating for channeling in the top encapsulation. Note that for thick hBN encapsulation, the modulation due to channeling in the hBN far exceeds that of the underlying TMD, making this process exceedingly challenging for increased hBN thickness.

### Monolayer TMD signal

Due to the three dimensional atomic structure of TMD layers, it is possible to measure the secondary yield modulation with azimuthal angle of a monolayer when imaging at a sufficiently shallow polar angle. Fig. S5a-c show SEM imaging, schematic and the azimuthal dependence for a $WSe_2$ monolayer on an $SiO_2$ substrate for a polar angle of 40°. These data indicate that SEM can be used to extract the lattice orientation of an exfoliated monolayer flake - information that is valuable for the assembly of twisted heterobilayer devices.

### Monolayer on bulk TMD signal

The ability to interrogate surface structure has the potential to elucidate more complex stacking orders, such as determining the relative rotation between a monolayer placed atop a bulk substrate. Fig. S6a shows an SEM image of a monolayer MoSe$_2$ flake stacked at a random angle on a thick multilayer MoSe$_2$ substrate (schematic in Fig. S6b). While the azimuthal rotation signal of the bulk (Fig. S6c) is relatively symmetric, the signal from the monolayer region displays a more complex pattern due to the partial channels between the monolayer and bulk. Taking the difference signal (Fig. S6d) reveals an asymmetric but roughly 60°-periodic structure. Further analysis may be able to use such a signal to determine the precise angular mismatch between the two layers.

**Computational modeling**

To compute primary electron transmission, atomic nuclei for the 2D material of interest are represented as points in 3D. Owing to the translational symmetry of the lattice, we consider incoming electrons across a hexagonal region corresponding to one lattice cell. A set of 10,000 to 250,000 coordinates within the hexagon is selected as entry points for the electron beams into the material, either at random with uniform probability or via a uniform grid. For each value of polar and azimuthal angle, a line through each point is considered, and the distance between the line and every lattice atom (i.e., the impact parameter, $b$ is calculated. A transmission probability of the form $1 - \frac{A}{(b/B)^2}$ is then taken, with the product over all lattice atoms giving the overall transmission probability for the line in question. Averaging over all lines then gives the overall transmission at the given $\theta$ and $\varphi$. Notably, though we only consider transmission across one lattice cell, the tails of the scattering probability mean that we must consider atoms outside of a one-cell-wide path. Typically, calculations were done considering atoms up to 4 inter-atomic distances away from the electron beam.

Transmission for the above figures was calculated using empirical values of $A = 0.1$ and $B = 0.18$Å. Computations for ABA/ABC graphene in Fig. 2 were performed for 5 layers stacked in ABCAB or ABABA orderings, using 100,000 lines within the lattice cell, with Monte Carlo coordinates. WSe$_2$ AB/BA computations were performed with 100,000 lines arranged on a grid, and verified with an equivalent model using 250,000 Monte Carlo generated lines. To represent the difference in atomic number in the WSe$_2$ lattice, an approximation is made whereby the tungsten atoms are modelled as two overlapping selenium atoms. Notably, the computed output that most closely resembles the data is at a polar angle of 43° rather than 40° at which the data was taken. This is most likely due to differences in interlayer spacing in AB/BA stacked TMD compared to the natural AA' stacking lattice constants used for the computation. The MoSe$_2$ computation in Fig. S2 is performed with a grid of 100,000 lines, at a polar angle of 27.5° and reproduces the periodicity but differs in some features. The difference could be attributed to the simplification of the multilayer to a model of only the top two layers, and also potential differences in secondary yield between molybdenum and selenium, along with effects of the 10nm gold encapsulation in attenuating secondary electron emission.

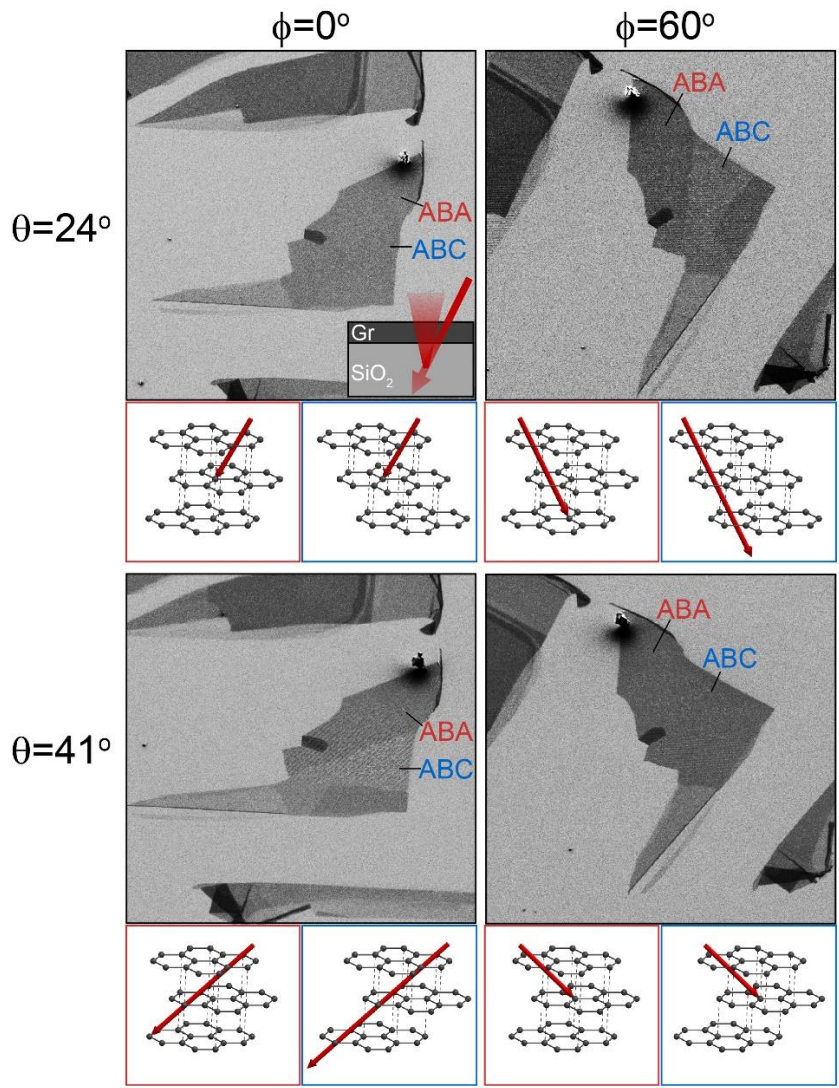

**Figure 1:** | **Polar angle variation. a-d)** SEM images of few-layer graphene containing ABA- and ABC-stacked regions, at different polar ($\theta = 24°, 41°$) and azimuthal ($\varphi = 0°, 60°$) angles. Schematics show channeling paths for ABA (red) and ABC (blue) stacking order at the different angle combinations. Since $\theta = 24(41)°$ corresponds to a shift of one (two) carbon-carbon bond lengths per layer, both polar angles enable complete channeling in the ABC region (causing bright SEM signal), but at different azimuthal angles.

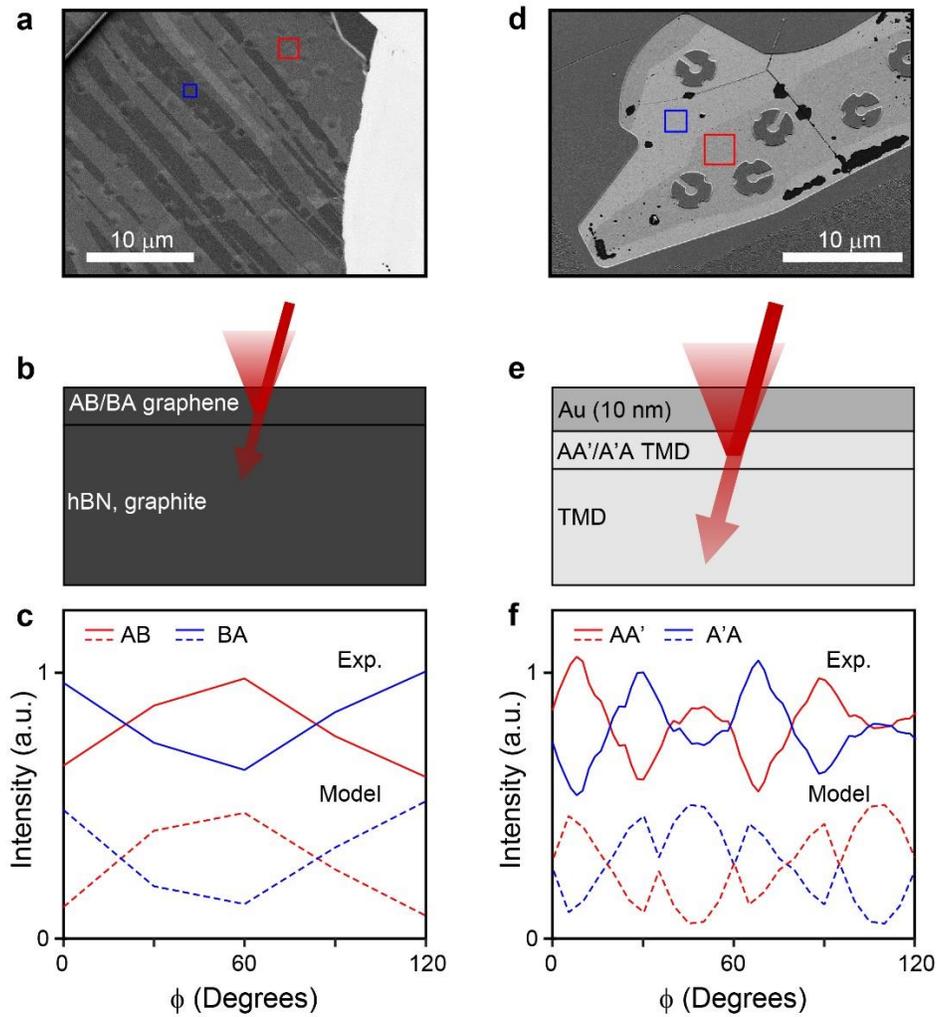

**Figure 2:** | **Imaging of stepped graphite and MoSe₂ a)** SEM image of stepped graphite flake using acceleration voltage of 500 eV and $\theta = 24°$. Alternating contrast is observed as the top two layers change between AB and BA stacking across the steps. **b)** Schematic of secondary generation, highlighting the sensitivity to the stacking order of the top few layers, which enables imaging of AB/BA domains on top of thick low-Z materials such as graphite or hBN. **c)** Azimuthal angle dependence of SEM signal in regions with AB (red) and BA (blue) surface stacking (solid lines). Dashed lines show predictions of model presented in main text. **d-f)** Same as **(a-c)**, but in stepped MoSe₂ underneath 0 nm Au. The more complex azimuthal dependence (**f**) is due to the 3D nature of the MoSe₂ layers, and also because the polar angle used here (30°) does not allow for maximal channeling.

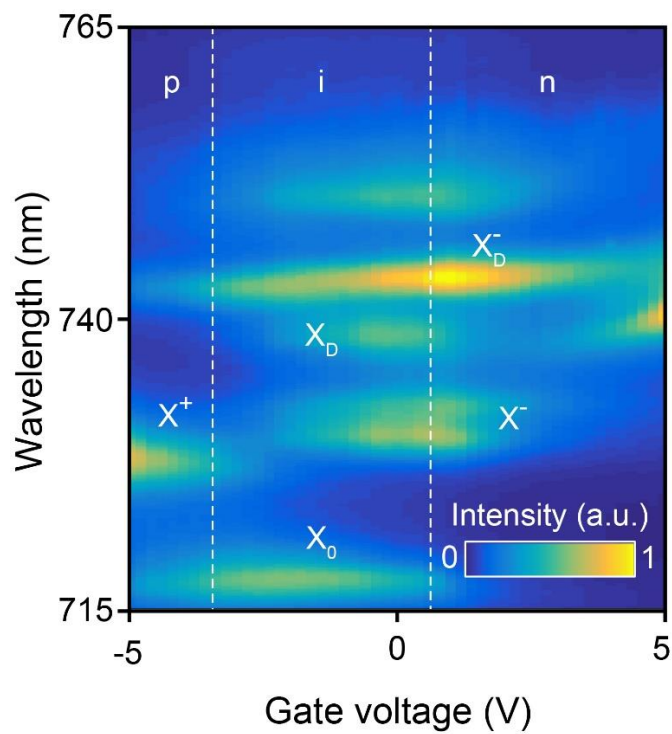

**Figure 3:** | **Compatibility of SEM technique with optoelectronic measurements** Gate-dependent photoluminescence spectra from hBN-encapsulated monolayer WSe$_2$, showing strong emission from both bright ($X_0$) and dark ($X_D$) excitons, with narrow linewidths (∼ 3 nm) and effective gate tunability. The latter is indicated by the appearance of charged excitons ($X^+$ and $X^-$).

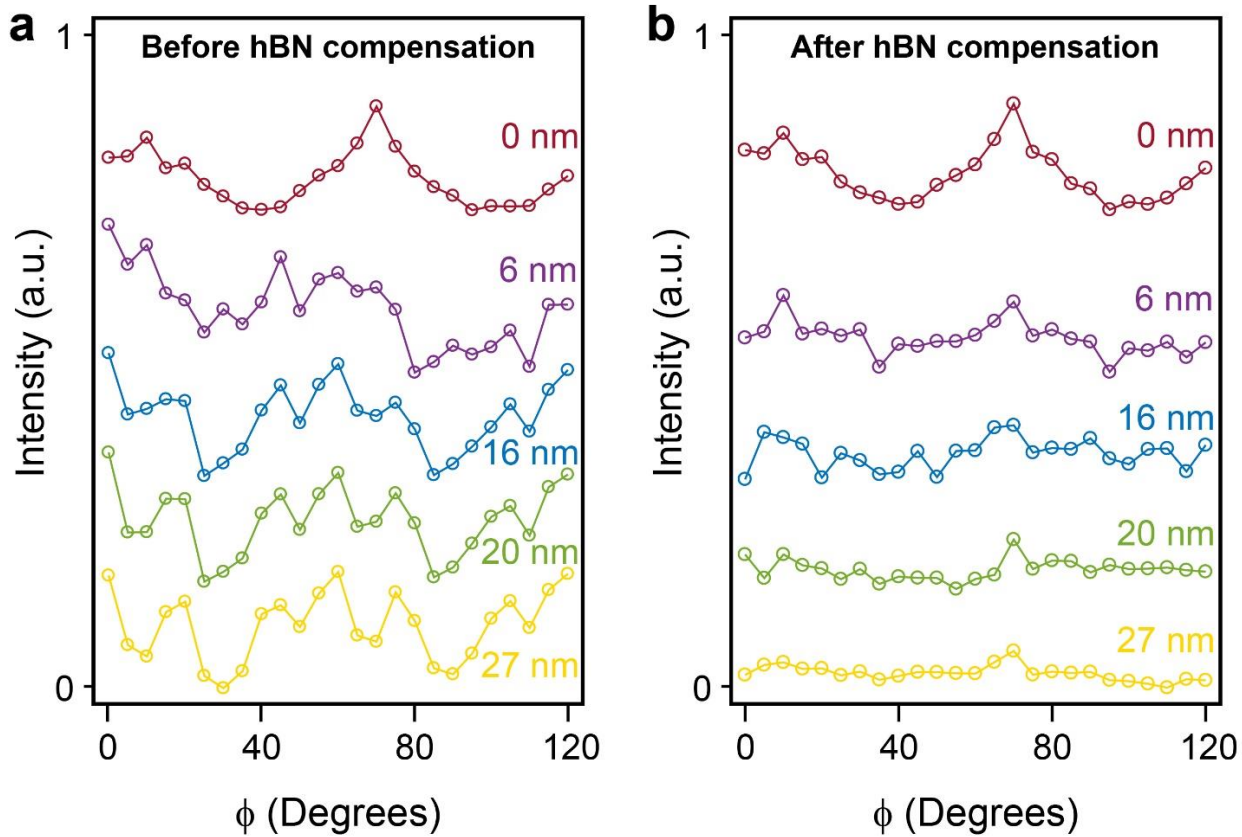

**Figure 4:** | **Compensation for channeling effects through top hBN a-b**) Azimuthal dependence of SEM signal from bi-layer TMD under hBN of varying thickness, before (**a**) and after (**b**) compensation for channeling effects in the hBN. We note that (**b**) is also shown in Fig. 4c in the main text.

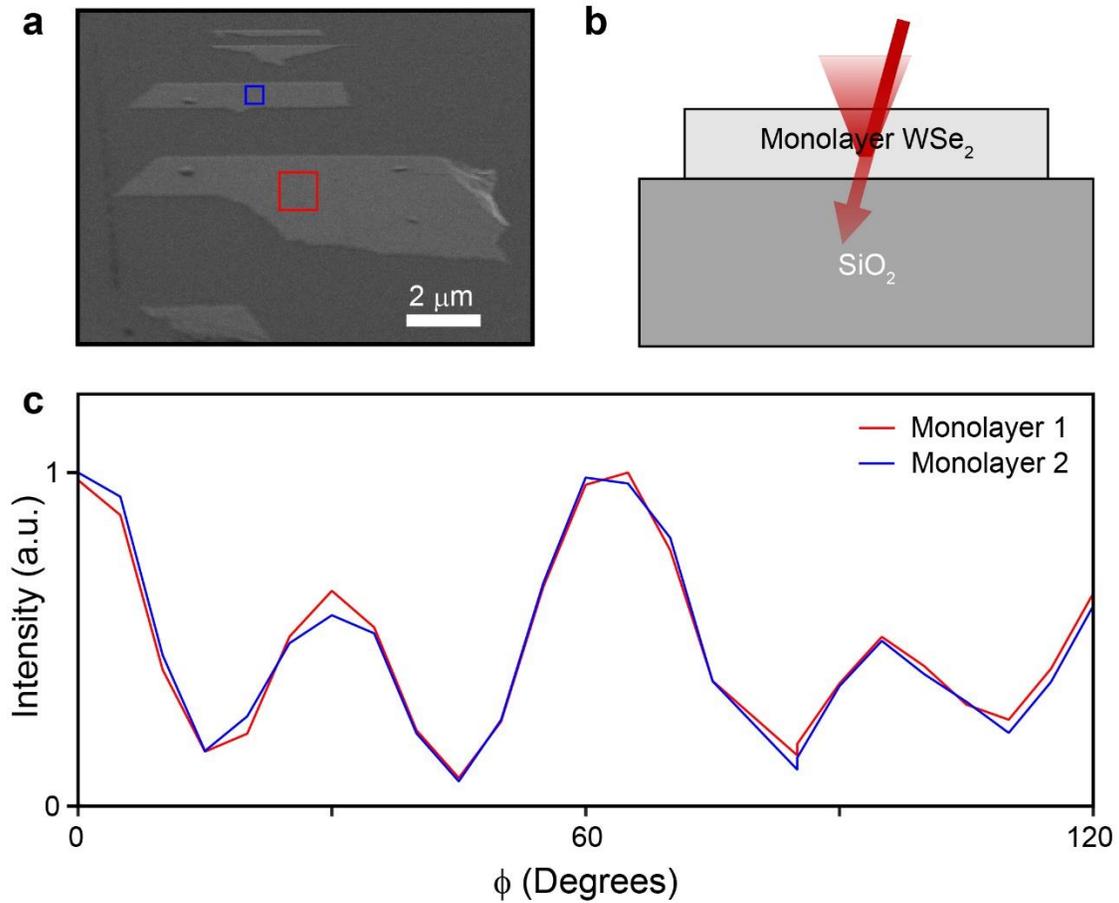

**Figure 5:** | **Imaging of monolayer TMD a**) SEM image of monolayer WSe₂ flakes on SiO₂ substrate, using acceleration voltage of 1 keV and $\theta = 40°$. **b)** Schematic of secondary emission in this system, showing stronger secondary signal from the TMD, due to its higher atomic mass. **c)** Azimuthal dependence of SEM signal from two separate monolayer WSe₂ flakes, indicated by boxes in (**a**). Due to the 3D structure of TMDs, amplitude modulation can be observed even from monolayers.

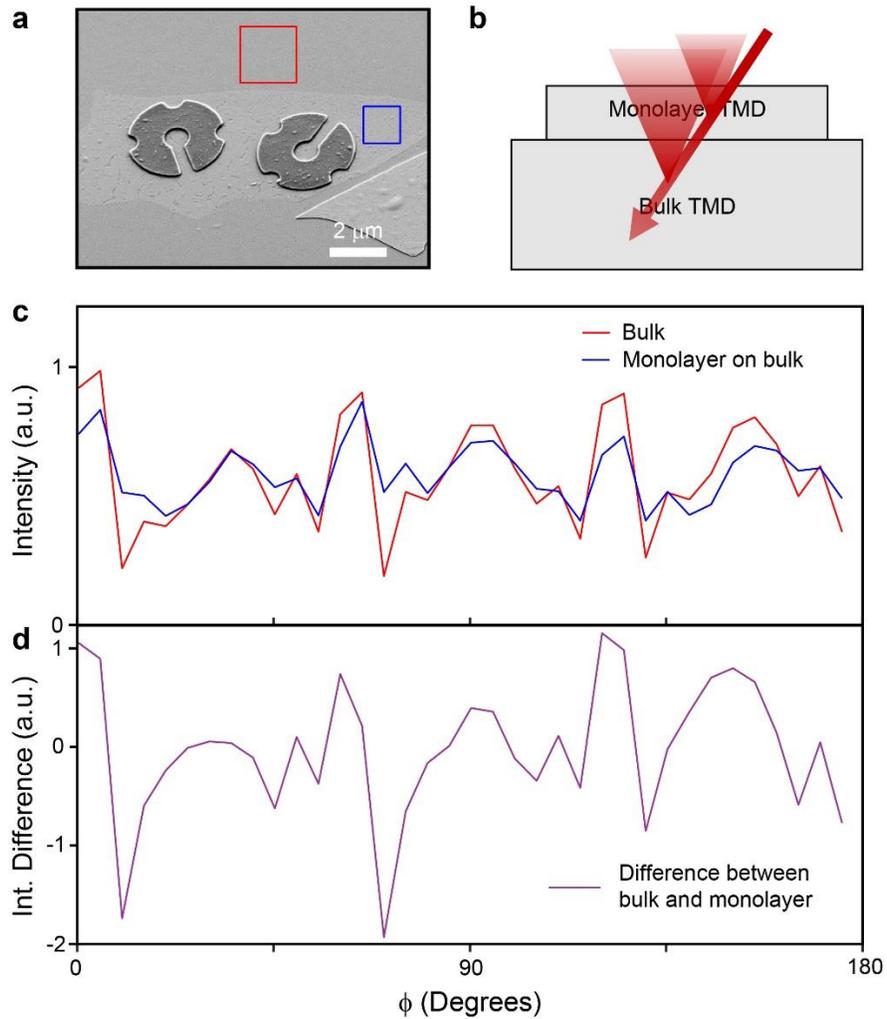

**Figure 6:** | **Imaging TMD monolayer on bulk a**) SEM image of monolayer $MoSe_2$ placed on top of a thick $MoSe_2$ flake. **b**) Schematic of secondary generation from both the monolayer and bulk flake. **c**) Azimuthal dependence of SEM signal from monolayer on bulk (blue) and only the bulk (red), collected from the boxed regions in (**a**). The more complex structure of the former is due to the partial channels of the combined system. **d**) Difference between the two curves in (**c**).

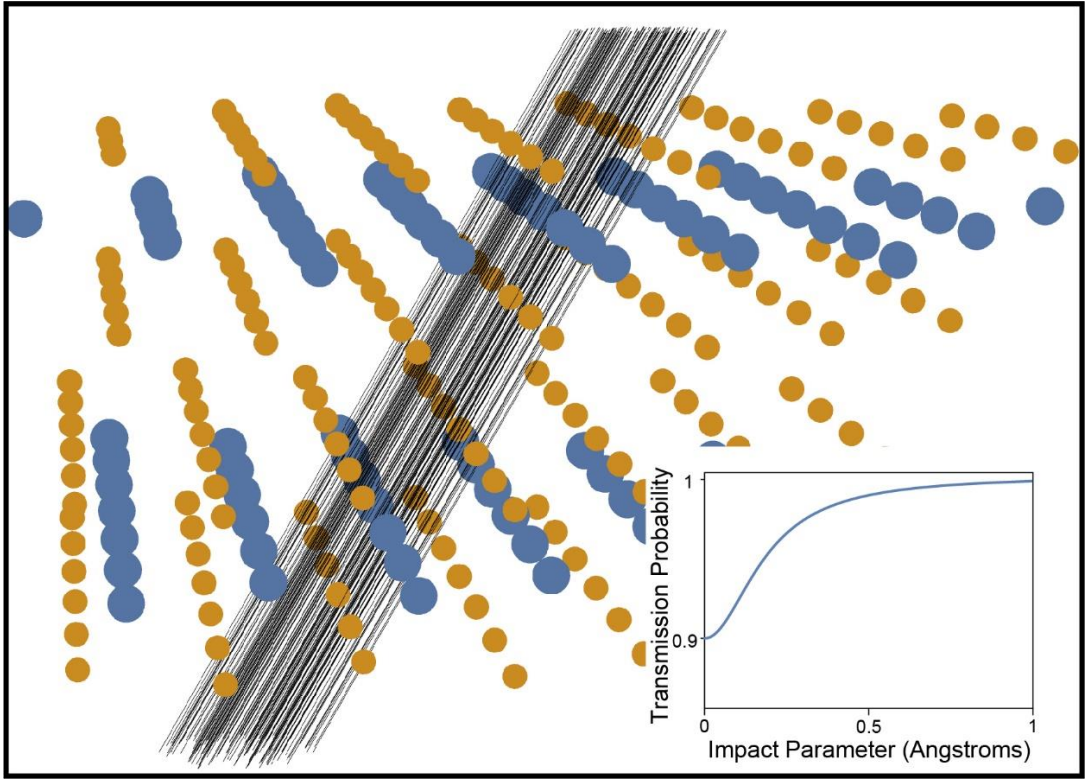

**Figure 7:** | **Computational modeling of SEM imaging** Schematic of computational method: Parallel primary beams (black lines) are positioned at Monte Carlo generated positions within an area corresponding to one unit cell, and their transmission probability through the lattice is calculated using a Lorentzian shaped scattering potential (inset) at each atom site.